\documentclass[twocolumn,aps,showpacs,groupedaddress]{revtex4}
\usepackage{array}
\usepackage{graphicx}

\def\be{\begin{equation}}
\def\ee{\end{equation}}
\def\bea{\begin{eqnarray}}
\def\eea{\end{eqnarray}}
\def\ba{\begin{array}}
\def\ea{\end{array}}
\def\bdm{\begin{displaymath}}
\def\edm{\end{displaymath}}

\begin{document}

\title{Mesoscopic p-wave superconductor near the phase transition temperature}

\author{Bor-Luen Huang and S.-K. Yip}

\affiliation{Institute of Physics, Academia Sinica, Taipei, Taiwan}

\date{\today }

\begin{abstract}
We study the finite-size and boundary effects on a p-wave superconductor
in a mesoscopic rectangular sample using Ginzburg-Landau and quasi-classical
Green's function theory.
Apart from a few very special cases, we find that
the ground state near the critical temperature always prefers a time-reversal
symmetric state, where the order parameter can be represented by a real vector.
For large aspect ratio,  this vector is parallel to the long side of
the rectangle.  Within a critical aspect ratio, it has instead a vortex-like structure,
vanishing at the sample center.

\end{abstract}


\maketitle

Studies of multicomponent superfluids and superconductors have excited many
for decades because of the diversity of textures, complex vortex
structures and collective modes.
The superfluid $^3$He with spin-triplet order parameter \cite{Leggett,VW} is
a well-established example. Many studies also show that superconductors with
multicomponent order parameters can also be found in, for example,
UPt$_3$ \cite{Sauls,JT} and Sr$_{2}$RuO$_{4}$ \cite{Maeno,MM}.
Recently, studies of multicomponent superconductor in a confined geometry
draw much attention due to advancements in nanofabrication.
Experiments claimed to find half-quantum vortices \cite{Budakian11}
and the Little-Parks effect \cite{Cai} in Sr$_{2}$RuO$_{4}$ quantum ring.
Surfaces are expected to have non-trivial effects on such superconductors.
Some theoretical works show that surface currents are present in broken
time-reversal symmetric superconductors.\cite{MS,SR04,Sauls2}
In considering Ru inclusions, Sigrist and his collaborators  \cite{Sigrist}
have shown that a time-reversal symmetry state can be favored near
the interface between Ru and Sr$_{2}$RuO$_{4}$ due to the boundary conditions.
In our previous work \cite{HY}, considering a thin circular disk with
smooth boundaries and applying Ginzburg-Landau (GL)  theory, we have shown
that a two-component p-wave superconductor can exhibit multiple
phase transitions in a confined geometry. At zero magnetic field,
the superconducting transition from the normal state was found to
be always first to a time-reversal symmetric state
(with an exception which occurs only far away from the
isotropic weak-coupling limit),
even though the bulk free energy may favor a broken time-reversal
symmetry state, which can exist at a lower temperature.
This time-reversal symmetric state has a vortex-like structure,
with order parameter vanishing at the center of the disk.
We have also argued there that these features are general,
do not rely on the GL approximation and should exist for also general geometries.
\cite{Vakaryuk}

In this paper, we investigate this question further
by considering rectangular and
square samples, employing both GL and quasiclassical
(QC) Green's function method.
Within GL, for rectangular samples with large aspect ratios,
we show that the phase transition
from the normal to the superconducting state is second-order and is
to a state with order parameter being a real vector
parallel to the long side of the sample.  For smaller aspect ratios,
the state near the transition temperature is again a time-reversal
symmetric state with a vortex at the center,
except for a square and only for gradient coefficients far
away from the weak-coupling limit,
much like what we found for the circular disk.
At not too small sizes, the results from QC
are qualitatively similar to GL except
for the critical sizes and aspect ratios obtained.
At very small sizes however, QC calculation suggests
that a more complicated situation can arise for some special
aspect ratios.  The transition can either become first-order,
or perhaps into a state with a more complicated order parameter.
In this paper, we shall mostly concentrate on the parameter region
where the phase transition is second-order and leave the detailed investigation
of the above mentioned special case to the future.

We shall thus consider a superconductor where its orbital part is
given by $\vec\eta=\eta_x\hat{x}+\eta_y\hat{y}$.  We shall consider
the dependences of $\eta_x$ and $\eta_y$ on the coordinates $x,y$,
assuming that they are constant along the $z$ direction.
We assume the length of the sample in x direction is $L$ and the width in y direction is $W$,
and these surfaces are smooth.  The effects of rough boundary
have been discussed in Ref.\cite{HY} for the circular disk.
We shall also limit ourselves to zero external magnetic fields.
Near the second-order transition temperature, the magnetic field generated by
the supercurrent is also negligible, hence
the vector potential can always be ignored.

First, we study this system via GL theory.
The GL free energy density per unit area for the bulk, $\mathcal{F}_b$, can be written as
\be
\mathcal{F}_b = \alpha (\vec \eta^* \cdot \vec \eta) + \ldots
\label{fb}
\ee
where $\alpha=\alpha'(t-1)$ with $\alpha'>0$, $t\equiv T/T_c^0$ is
the ratio of the temperature $T$ relative to the bulk transition temperature $T_c^0$,
and  $\ldots$ represents terms higher power in the order parameter
which are irrelevant below since we are interested only in the physics
at the (modified) transition temperature $T_c$.
In the presence of spatial variations,  there is an additional contribution to the free energy
given by
\be
\ba{ll}
\mathcal{F}_g  & = K_1 (\partial_j \eta_l) (\partial_j \eta_l)^*
    + K_2 (\partial_j \eta_j) (\partial_l \eta_l)^* \\
  & + K_3 (\partial_j \eta_l) (\partial_l \eta_j)^*
    + K_4[(\partial_x\eta_x)^2+(\partial_y\eta_y)^2],
   \label{fg}
\ea
\ee
where repeated indices $j,l$ in the first three terms are summed over $x,y$,
and the last term describes crystal anisotropy.\cite{compare-A}
Within weak coupling approximation, particle-hole symmetry, and for an
isotropic Fermi surface, $K_1=K_2=K_3 > 0$ and $K_4=0$, but we shall
treat these coefficients as general parameters.

The GL equations need to be accompanied by
boundary conditions.
The perpendicular component of the
order parameter at the surface should vanish \cite{AGD}.
Thus, for a point at the surface where the normal is $\hat n$,
$\hat n \cdot \vec \eta = 0$.
For a smooth surface,
the parallel component $\eta_{\|}$ should have vanishing normal gradient \cite{AGD}.
That is, at the surface, $(\hat n \cdot \nabla) \eta_{\|} = 0$.

GL equations for $\eta_{x,y}$ can be obtained by the variation principle.
Near the critical temperature, we can linearize these equations.
The easiest way to match the boundary conditions is to superimpose the Fourier components.
Written in matrix form, the GL equations
for the Fourier component $\vec q$ become
\bea
 & &
  \left(
  \ba{cc}
   K_1 q^2 + K_{234}q_x^2  &  K_{23}q_x q_y \\
  K_{23}q_x q_y & K_1 q^2 + K_{234}q_y^2
  \ea\right)
  \left(
  \ba{c}
  \eta_{x,\vec q} \\ \eta_{y, \vec q}
  \ea\right)    \nonumber \\
  &  &  \qquad  \qquad   = \alpha' (1 - t)
    \left(
  \ba{c}
  \eta_{x, \vec q} \\ \eta_{y, \vec q}
  \ea\right).
  \label{e2}
\eea
Here, $\vec q$ is the wavevector, $q^2=q_x^2+q_y^2$, $K_{23}=K_2+K_3$, and
$K_{234}=K_2+K_3+K_4$.

It is easy to find that, if $q_x=0$ or $q_y=0$, Eq.(\ref{e2}) decouples.
We obtain either (i) $\eta_{x,\vec q}\ne0$ with $\eta_{y, \vec q}=0$ or
 (ii) $\eta_{y,\vec q}\ne0$ with $\eta_{x,\vec q}=0$.
 We call these solutions as A phases.
In case (i), we have two possibilities. One is $\hat{q}=\hat{x}$ and
the critical temperature is determined by $\alpha' (1-t)=K_{1234}q^2$.
Because $\eta_x(x)$ is independent of $y$, the possible solutions are
$\eta_x=X\sin\frac{m\pi x}{L}$, satisfying the boundary conditions, $\eta_x=0$,
at $x=0$ and $L$. Here $X$ is a constant and $m$ is an integer. The best choice
is $m=1$ and the critical temperature is $\alpha' (1-t)=K_{1234}(\pi/L)^2$.
We call this the $A_1$ phase.
The other is $\hat{q}=\hat{y}$. The order parameter $\eta_x(y)$ is
independent of $x$. Thus it is not possible to satisfy the boundary conditions at
$x=0$ and $L$. In case (ii), the best solution is $\eta_y\propto\sin\frac{\pi y}{W}$,
which is just the solution in case (i) with $x\leftrightarrow y$. The critical
temperature is determined by $\alpha'(1-t)=K_{1234}(\pi/W)^2$. We call this the $A_2$ phase.

If both $q_x$ and $q_y$ are non-zero, both $\eta_{x,\vec q}$, $\eta_{y,\vec q}$ are finite.
 We define this kind of solution as B phase.
To simplify the calculations, we ignore crystal anisotropy for
the moment, and set $K_4 = 0$.
From Eq.(\ref{e2}), we find the smallest eigenvalue is
$K_1 q^2$ (for $K_{23}>0$).
To have the normal component  to the surfaces at $x = 0$ and $L$ to vanish,
$\eta_x$ must have the factor $\sin(m\pi x/L)$, where $m$ is an integer.
Because the boundary conditions $\partial\eta_x/\partial y=0$ at $y=0$ and $W$,
$\eta_x$ should be proportional to $\cos(n\pi y/W)$, with $n$ also an integer.
 We can use the same arguments
for $\eta_y$. Therefore
\be
  \ba{l}
  \eta_x=X \sin\frac{m\pi x}{L}\cos\frac{n\pi y}{W}, \\
  \eta_y=Y \cos\frac{m\pi x}{L}\sin\frac{n\pi y}{W}.
  \ea
  \label{e3}
\ee
The critical temperature is highest for $m=n=1$, and thus
determined by $\alpha'(1-t)=K_{1}[(\pi/L)^2+(\pi/W)^2]$.
In order to satisfy Eq.(\ref{e2}), we need $(\pi/L)X+(\pi/W)Y=0$, which means $X$ and
$Y$ are relatively real and have opposite signs.

\begin{figure}
\begin{center}
\rotatebox{0}{\includegraphics*[width=80mm]{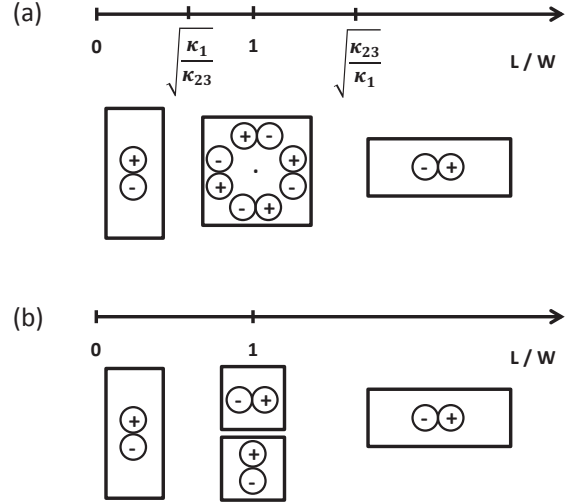}}
\caption{GL phase diagram at the transition temperature.  $K_4=0$.
(a) $K_{23}/K_1 > 1$ (b) $K_{23}/K_1 < 1$.
}\label{fig1}
\end{center}
\end{figure}

\begin{figure}
\begin{center}
\rotatebox{0}{\includegraphics*[width=75mm]{fig2a.eps}}
\rotatebox{0}{\includegraphics*[width=80mm]{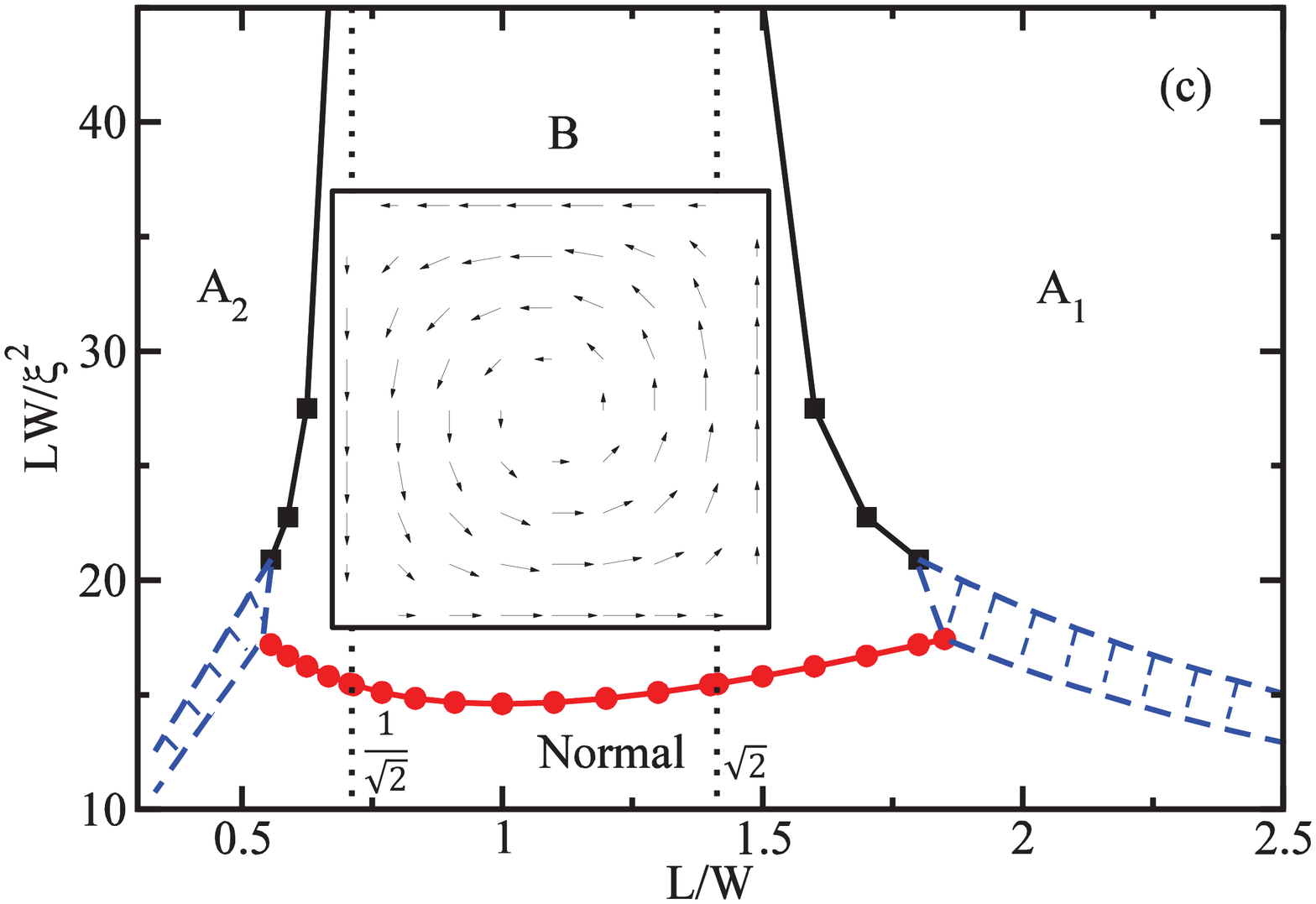}}
\caption{(a) Critical temperatures from GL  and QC theories for square samples.
(b) Critical temperatures in QC theory for different $\rho$'s. As the size of the system is reduced,
the ground state near the critical temperature can change from A to B.
The possible first-order phase transition is indicated in the region between dash lines.
(c) Phase diagram for different sizes. 
The white regions correspond to second-order normal to superconducting transitions;
the dashed regions indicate possible first-order phase
transitions.
Inset: Order parameter for a square sample.
Note that all results are for $K_1=K_2=K_3$. $\xi$ is the coherence length.
}\label{fig2}
\end{center}
\end{figure}

Comparing the transition temperatures of the
A and B phases, we find that the system prefers the $A_1$ phase for $L\gg W$.
For $L \sim W$, it prefers the B phase. We define the aspect ratio of
the sample as $\rho=L/W$. The critical aspect ratio separating these two phases is
\be
  \rho_c=\sqrt{{K_{23}}/{K_1}}.
  \label{cr}
\ee
With the same reasoning, the system is in $A_2$ phase for $W\gg L$ but prefers the
B phase if $\rho$ is larger than $\rho_c^{-1} = \sqrt{K_1/K_{23}}$.
The phase diagram is shown in Fig.\ref{fig1}(a). The cartoon pictures under
the ruler are used to specify main characteristic of the order parameter
in the corresponding phases.
The solution for the middle case (B phase) is qualitatively same as the circular disk
in \cite{HY}.
Because the relatively real coefficients in Eq.(\ref{e3}),
we can represent $\vec{\eta}$ by a real vector,
as done in the inset of Fig.\ref{fig2}(c).
It is clear that the order parameter forms a vortex-like structure,
and vanishes at the sample center.

When crystal anisotropy is included, the critical ratio becomes
\be
\rho_c=\sqrt{\frac{K_{1234}K_{234}-K_4(K_{23}+K_{234})}{K_1 K_{1234}}}.
\ee
It reduces to Eq.(\ref{cr}) for $K_4=0$. For small $K_4$,
the critical ratio is $\sqrt{2-K_4/K_{123}}$ within the weak-coupling limit.
It shows that the crystal anisotropy for $K_4>(<)0$ stabilizes(destabilizes) the order parameter
with the direction parallel to the long side of the sample.
The phase diagram is similar to Fig.\ref{fig1}(a) except smaller(larger) region for
vortex state. We ignore
crystal isotropy in the following.

With decreasing $K_{23}/K_1$, the stability region for B phase becomes narrower.
This phase diagram (Fig.\ref{fig1}(a)) will change qualitatively if $K_1>K_{23}$
The $B$ phase is
never stable (at $T_c$), and the system is in the $A_1$ phase if $\rho>1$,
and in the $A_2$ phase if $\rho< 1$.  The square sample $\rho=1$ forms a special case,
where the system still has $C_4$ symmetry in real space,
and the $A_1$ and $A_2$ phases are therefore degenerate. One can combine these
two solutions with a phase difference. As a result,
if the higher order terms in eq (\ref{fb}) for the bulk free energy
prefer a time-reversal-symmetry-broken state,
then the system would enter such a state directly at $T_c$.
Therefore, the phase diagram becomes
Fig.\ref{fig1}(b) for $K_{23}<K_1$, where the ground state at
$\rho=1$ should break the time-reversal symmetry. \cite{footnote2}

Our results for the square here provide further understanding of those we obtained earlier for
the circular disk \cite{HY}.  There, we found that, for $K_{1,2,3}$ near the
weak-coupling values ($K_1=K_2=K_3$) the phase transition from the
normal state is always to the state (named $n=1$ there) which preserves
time-reversal but with a vortex at the center.  That phase obviously has
the same qualitative behavior as our B phase here.   For sufficiently
small $K_{23}/K_1$, we found that the system can enter a broken time-reversal
symmetry state directly.  We find the same results here though
the critical value for $K_{23}/K_1$ obviously can depend on the geometry.

In order to check the validity of the phase diagram from GL theory, we employ
quasi-classical (QC) Green's function theory.
For simplification, we focus on the isotropic and weak-coupling case.
As shown in Ref.\cite{HY}, we have, after linearizing in the order parameter,
\be
  ( 2 i \epsilon_n + i v_f \hat p \cdot \vec \nabla) f
  = 2 i \pi ( {\rm sgn} \epsilon_n ) \Delta,
 \label{qcf}
\ee
where $f (\hat p,  \epsilon_n, \vec r)$ and $\Delta (\hat p, \vec r)$
describe separately the off-diagonal parts of the
QC propagator and pairing function, $\hat{p}$ is the momentum direction,
$\epsilon_n$ is Matsubara frequency, and $v_f$ is the Fermi velocity.
With pairing interaction written as $V_1 \hat p \cdot \hat p'$,
the gap  equation reads
\be
  \Delta (\hat p, \vec r) = N(0) T V_1 \sum_{n}
  < ( \hat p \cdot \hat p') f (\hat p', \vec r, \epsilon_n) >
  \label{gapeqn}
\ee
where the angular bracket denotes angular average over $\hat p'$ and $N(0)$ is the
density of states at the Fermi level.
For our square geometry and assuming smooth surfaces,
we have the
boundary conditions $f(\theta)=f(\pi-\theta)$ at x=0 and L
and $f(\theta)=f(-\theta)$ at y=0 and W.
Here $\theta$ is the angle between $\hat{p}$ and $\hat{x}$.

Before solving the case in a confined rectangle, we like to mention the connection between
GL theory and QC theory for the bulk. To zeroth order in gradient,
one finds
\be
1=\pi N(0) V_1 T_c^0 \sum_{\epsilon_n}\frac{1}{2|\epsilon_n|} ,
\label{zeroth-order}
\ee
which defines the bulk transition temperature $T_c^0$.
The first order for $f$ is odd in $\epsilon_n$ and will not contribute to the gap equation.
In the second order, we recover the GL theory with
\be
  \frac{K_1}{\alpha'}=2\pi T_c^0\sum_{\epsilon_n}\frac{v_f^2}{4|\epsilon_n|^3}
  <\cos^2\theta\sin^2\theta>,
  \label{compare}
\ee
and similar expressions for $K_{2,3}$, with $K_1=K_2=K_3$.
Our equations are consistent with those in Ref. \cite{Leggett,VW,Agterberg}

For the $A_1$ phase,
we shall show that we can have
a self-consistent solution in QC theory with
\be
  \Delta(\hat{p},\vec{r})=X\sin\frac{\pi x}{L}\cos\theta ,
  \label{eq-da}
\ee
the order parameter suggested by the GL theory.
One finds that  $f$ is independent of $y$. With the ansatz
\be
  f(\theta,\epsilon_n;x)=C_1(\theta,\epsilon_n)\sin\frac{\pi x}{L}
  + C_2(\theta,\epsilon_n)\cos\frac{\pi x}{L}
  \label{eq-fa}
\ee
satisfying the boundary conditions, solving for $C_{1,2}(\theta,\epsilon_n)$
via eq.(\ref{qcf}) and using (\ref{zeroth-order}), we find
\be
  \mbox{ln}\frac{T_c^0}{T_c}=2\pi T_c\sum_{\epsilon_n=-\infty}^{\infty}\langle
  \frac{(\frac{v_f\pi}{L})^2\cos^4\theta}{4|\epsilon_n|^3
  [1+\frac{(\frac{v_f\pi}{L}\cos\theta)^2}{4\epsilon_n^2}]}\rangle.
  \label{ta}
\ee
For large $L$, one can replace the bracket in the denominator by $1$, the LHS by $(1-t)$,
recovering the GL result using Eq.(\ref{compare}).
Hence we see that, beyond GL, one needs simply to include extra factors in the
denominator of Eq.(\ref{ta}) and include the ln on the LHS.

Now, we consider the B phase. The order parameter is
\be
  \Delta(\hat{p},\vec{r})=X\sin\frac{\pi x}{L}\cos\frac{\pi y}{W}\cos\theta+
  Y\cos\frac{\pi x}{L}\sin\frac{\pi y}{W}\sin\theta.
  \label{eq-db}
\ee
From previous experience, it suggests that the solution has the following form
\be
\ba{ll}
  f= & C_1(\theta,\epsilon_n)\sin\frac{\pi x}{L}\cos\frac{\pi y}{W}+
  C_2(\theta,\epsilon_n)\cos\frac{\pi x}{L}\sin\frac{\pi y}{W}+ \\
  & C_3(\theta,\epsilon_n)\cos\frac{\pi x}{L}\cos\frac{\pi y}{W}+
  C_4(\theta,\epsilon_n)\sin\frac{\pi x}{L}\sin\frac{\pi y}{W}.
\ea
  \label{eq-fb}
\ee
To simplify writing, let $A=\pi v_f/L$, $B=\pi v_f/W$. Again solving
for $f$ from (\ref{qcf}), we obtain the following coupled linear equations in $X,Y$:
\bea
  X\mbox{ln}\frac{T_c^0}{T_c}&=&2\pi T_c\sum_{\epsilon_n}\langle
  \frac{[(c_1+c_2)\cos^2\theta] X+c_3Y}{|\epsilon_n| D}\rangle.
  \label{tb1} \\
  Y\mbox{ln}\frac{T_c^0}{T_c}&=&2\pi T_c\sum_{\epsilon_n}\langle
  \frac{[(c_1+c_2)\sin^2\theta] Y+c_3 X}{|\epsilon_n| D}\rangle.
  \label{tb2}
\eea
Here $c_1 = (A^2\cos^2\theta+B^2\sin^2\theta)/(4\epsilon_n^2)$,
$c_2 = (A^2\cos^2\theta-B^2\sin^2\theta)^2/(4\epsilon_n^2)^2$,
$c_3 = (AB\sin^2\theta\cos^2\theta)/(2\epsilon_n^2)$, and $D=1+2c_1+c_2$.
The critical temperature of the B phase is determined by the point
which allows non-trivial $X$ and $Y$.
We note that if one keeps only the lowest orders in $A^2$ and $B^2$, $D\rightarrow 1$,
and replaces the ln's on LHS by $(1-t)$,
then Eq.(\ref{tb1}) and Eq.(\ref{tb2}) recover the corresponding equations (\ref{e2})
in GL theory.

In Fig.\ref{fig2}(a), we compare the critical temperatures for different sizes of
square samples between GL and QC theories. We use the coherence length
$\xi \equiv \sqrt{K_{123}/\alpha'}$ as the unit for length ($\xi = 0.199909 v_f /T_c^0$
for QC). In GL theory, $(1-t)$, the relative suppression of critical temperature,
is inversely proportional to the square of the length scale of the system.
Therefore it is more convenient to set the vertical axes of the phase diagram
to be $(\xi/L)^2$. We obtain the straight line with crosses for the critical temperature
of A phase and the line with pluses for that of B phase. It shows that the system prefers the B
phase. On the other hand, we also present the critical temperatures calculated from QC theory.
The line with squares is for the B phase and the line with circles is for the A phase.
As expected, it shows that the results from QC theory are consistent with those from GL theory
near $t=1$, corresponding to $\xi/L \ll 1$.
In addition to the fact that the B phase is still preferred,
we see that the critical temperature is more suppressed than GL
as the size of the system is smaller. \cite{cp}

As the aspect ratio $\rho$ becomes larger than 1, GL theory shows that there is
a phase transition from the B phase to the A phase as $\rho$ increases beyond the
critical ratio $\sqrt{K_{23}/K_1}$ (Fig.1(a)). In Fig.2(b),
we find that this critical ratio ($\sqrt{2}$ here) still applies for large systems.
However, we find that the B phase occupies a slightly larger $\rho$ region when the size decreases.
An example is in Fig.2(b), where we show
that the system processes a phase transition from the A phase (thick black line) to the B phase
for $\rho=1.6$ (thin red line) around $t=0.67$ as the size of the system becomes smaller.
As the aspect ratio increases further, the situation becomes more complicated because the phase
boundary for A phase is not a monotonic function in $t$.
The shape of the curve suggests that, at lower  temperatures,
the transition from the normal to the superconducting state cannot
be a second-order transition to the A$_1$ phase as described here.
One possibility is a first-order phase transition between
the normal and the superconducting A$_1$ phase, analogous to \cite{Sarma}.
However, transition into a more exotic order parameter  structure cannot
be ruled out \cite{FFLO}.
We shall leave the detailed investigation of
this question
for the future.
For illustrative purposes, we indicate the resulting phase diagram by postulating a
first-order transition line somewhere between the two dashed lines in Fig.2(b).
Our obtained phase diagram for the normal to superconducting
transition, with instead now the sample area as
the vertical axes, is as shown in Fig.2(c).
For the square samples, the stable superconducting phase is the B phase, with a vortex-like structure
at the center, similar to the $n=1$ state in \cite{HY}
obtained in the disk geometry.

In conclusion, we studied
a two-component p-wave superconductor
in a rectangular geometry near its transition temperature.
The order parameter can behave differently depending on the aspect ratio and size.
Except for some special regions in parameter space,
the phase always preserves time-reversal symmetry.
Our results give further support to those obtained in \cite{HY}.

This work is supported by the National Science Council of Taiwan
under grant number NSC 101-2112-M-001 -021 -MY3.


\begin{thebibliography}{plain}
\bibitem{compare-A} Eq.(\ref{fg}) can also be rewritten as
$\{K'_1|\partial_x\eta_x|^2 + K'_2|\partial_y\eta_x|^2
+ K'_3(\partial_x\eta_y)(\partial_y\eta_x)^* + K'_4(\partial_y\eta_x)(\partial_x\eta_y)^*\}
+ \{x\leftrightarrow y\}$,
the same form as in Ref.{\cite{Agterberg}}
with $K'_1=K_{1234}$, $K'_2=K_1$, $K'_3=K_2$, and $K'_4=K_3$.
\bibitem{footnote2} On the other hand, if the higher order terms in eq (\ref{fb})
favor a time-reversal symmetric state for the bulk, then the system at
$\rho=1$ would then be either in the $A_1$ or $A_2$ phase,
thus spontaneously breaking the $C_4$ rotational symmetry.
We thus believe that the speculation in our footnote 27 in ref \cite{HY} is
probably incorrect.
\bibitem{cp} In our previous study \cite{HY} in circular disks, we found
instead that the critical temperature is independent of the radius $R$. We
speculate that the present suppression of $T_c$ is due to the existence of corners.
\bibitem{FFLO} We note that our A$_1$ phase has order parameter dependent on
$x$ but not $y$.  Since the surfaces at $y = 0$ and $W$ are taken to be
perfectly reflecting, formally our solution is the same as an infinitely
long strip along the $y$ direction, with the order parameter vector forced
to be along the $x$-axis.  In this case, similar arguments as in
\cite{Vorontsov} shows that, for this infinitely long strip, the system
may prefer an Fulde-Ferrell-Larkin-Ovchinnikov like structure along the $y$-direction,
with the relevant wave-vector of the order of $L^{-1}$.  Our system instead
has finite extent $W$ along the $y$ direction.  It remains to be investigated
whether an analogous phase can occur in our system, though we suspect
that this is somewhat unlikely if $W$ is much less than $L$.
\end{thebibliography}
\end{document}